\documentclass[aoas,nameyear,seceqn,dvips]{arximspdf}
\usepackage{dcolumn,multirow}
\usepackage{graphicx}

\doi{10.1214/10-AOAS352}
\volume{4}
\issue{4}
\pubyear{2010}
\firstpage{1913}
\lastpage{1941}

\makeatletter
\newproclaim{assumption}{Assumption}
\newtheorem{theorem}{Theorem}[section]
\newcolumntype{d}[1]{D{.}{.}{#1}}
\newtheorem{lemma}[theorem]{Lemma}
\newtheorem{proposition}[theorem]{Proposition}
\makeatother

\begin{document}
\begin{frontmatter}

\title{Nonparametric inference of doubly stochastic Poisson process
data via the kernel method\thanksref{TX1}}
\runtitle{Kernel inference of Cox process data}

\thankstext{TX1}{Supported in part by the NSF Grant DMS-04-49204 and the NIH/NIGMS Grant
1R01GM090202-01.}

\begin{aug}
\author[A]{\fnms{Tingting} \snm{Zhang}\ead[label=e1]{tz3b@virginia.edu}}
\and
\author[B]{\fnms{S. C.} \snm{Kou}\ead[label=e2]{kou@stat.harvard.edu}\corref{}}
\runauthor{T. Zhang and S. C. Kou}
\affiliation{University of Virginia and Harvard University}
\address[A]{Department of Statistics\\
University of Virginia\\
Charlottesville\\
Virginia 22904\\
USA\\\printead{e1} }              
\address[B]{Department of Statistics\\
Harward University\\
Cambridge, Massachusetts\\
USA\\\printead{e2} }
\end{aug}

\received{\smonth{6} \syear{2009}}
\revised{\smonth{4} \syear{2010}}

\begin{abstract}
Doubly stochastic Poisson processes, also known as the Cox processes,
frequently occur in various scientific fields. In this article, motivated
primarily by analyzing Cox process data in biophysics, we propose a
nonparametric kernel-based inference method. We conduct a detailed study,
including an asymptotic analysis, of the proposed method, and provide
guidelines for its practical use, introducing a fast and stable regression
method for bandwidth selection. We apply our method to real photon arrival
data from recent single-molecule biophysical experiments, investigating
proteins' conformational dynamics. Our result shows that conformational
fluctuation is widely present in protein systems, and that the fluctuation
covers a broad range of time scales, highlighting the dynamic and complex
nature of proteins' structure.
\end{abstract}


\end{frontmatter}

\section{Introduction}

Poisson processes, fundamental to statistics and probability, have wide
ranging applications in sciences and engineering. A special class of Poisson
processes that researchers across different fields frequently encounter is
the doubly stochastic Poisson process. Compared to the standard Poisson
process, a key feature of a doubly stochastic one is that its arrival rate
is also stochastic. In other words, if we let $N(t)$ denote the process and
let $\lambda (t)$ denote the arrival rate, then, conditioning on $\lambda (t)$,
\[
N(t)|\lambda (t)\sim \mbox{inhomogeneous Poisson process with rate } \lambda(t),
\]
where $\lambda (t)$ itself is a stochastic process [\citet{CoxIsh1980};
\citet{DalVer1988}; \citet{Karr1991}; \citet{KarTay1981}]. In the
literature such processes are also referred to as Cox processes in honor of
their discoverer [Cox (\citeyear{Cox1955a}, \citeyear{Cox1955b})].

We consider the inference of Cox processes with large arrival rates in this
article. Our study is primarily motivated by the frequent occurrences of Cox
process data in biophysics and physical chemistry. In these fields,
experimentalists commonly use fluorescence techniques to probe a biological
system of interest [\citet{KriBon2002}], where the system is placed
under a laser beam, and the laser excites the system to emit photons. The
experimental data consist of photon arrival times with the arrival rate
depending on the \textit{stochastic} dynamics of the system under study (for
example, the active and inactive states of an enzyme can have different
photon emission intensities). By analyzing the photon arrival data, one aims
to learn the system's biological properties, such as conformational dynamics
and reaction rates.

Although we mainly focus on biophysical applications, we note that Cox
processes also appear in other fields. In neuroscience Cox process data
arise in the form of neural spike trains---a chain of action potentials
emitted by a single neuron over a period of time [\citet{GersKis2002}]---from which
researchers seek to understand what information is conveyed in
such a pattern of pulses, what code is used by neurons to transmit
information, and how other neurons decode the signals, etc.
[\citet{Bialek1991}; \citet{Barbieri2005}; \citet{Rieke1996}].
Astrophysics is another area where Cox process data often occur. For
example, gamma-ray burst signals, pulsar arrival times and arrivals of
high-energy photons [\citet{Meegan1992}; \citet{Scargle1998}] are studied
to gain information about the position and motion of stars relative to the
background [\citet{CarOst2007}].

Previous statistical studies of Cox process data in the biophysics and
chemistry literature mainly focus on constructing/analyzing parametric
models. For instance, continuous-time Markov chains and stationary Gaussian
processes have been used to model the arrival rate $\lambda (t)$ for
enzymatic reactions [\citet{English2006}; \citet{Kou2005b}; \citet{Kou2008b}], DNA dynamics [\citet{KouXie2005a}], and proteins'
conformational fluctuation [\citet{Min2005b}; \citet{KouXie2004}; \citet{Kou2008a}].

Although effective for studying the stochastic dynamics of interest when
they are correctly specified, parametric models are not always applicable
for data analysis, especially when researchers (i) are in the early
exploration of a new phenomenon, or (ii) are uncertain about the correctness
of existing models, and try to avoid drawing erroneous conclusions from
misspecified parametric models. Owing to its flexibility and the intuitive
appeal of ``learning directly'' from data,
we focus on the \textit{nonparametric} inference of Cox process data in this
paper. In particular, we develop kernel based estimators for the arrival
rate $\lambda (t)$ and its autocorrelation function (ACF). The ACF is of
interest because it directly measures the strength of dependence and reveals
the internal structure of the system. For example, for biophysical data, a
fast decay of the ACF, such as an exponential decay, indicates that the
underlying biological process is Markovian and that the biomolecule under
study has a relatively simple conformation dynamic, whereas a slow decay of
ACF, such as a power-law decay, signifies a complicated process and points
to an intricate internal structure/conformational dynamic of the
biomolecule. Thus, in addition to discovering important characteristics
of the stochastic dynamics under study, the autocorrelation function can
also be used to test the validity of parametric models.

Kernel smoothing and density estimates have been extensively developed in
the last three decades; see, for example, \citet{Silverman1986}, \citet{Eubank1988}, \citet{Muller1988},
\citet{Hardle1990}, \citet{Scott1992}, \citet{Wahba1990}, \citet{WanJon1994}, \citet{FanGij1996}
 and \citet{BowAzz1997}. Meanwhile,
kernel estimators of spatial point processes motivated by applications in
epidemiology, ecology and environment studies have been proposed; see Diggle
(\citeyear{Diggle1985}, \citeyear{Diggle2003}), \citet{StoSto1994}, \citet{MolWaa2003},
Guan, Sherman and Calvin~(\citeyear{GuanShe2004}, \citeyear{GuanShe2006}) and \citet{Guan2007}. Compared to these
spatial applications, the Cox processes that we encounter in biophysics have
some unique features: (i) the arrival rates are usually large because strong
light sources, such as laser, are often used; (ii) the data size tends to be
large, since one can often control the experimental duration; (iii) both
short-range and long-range dependent processes can govern the underlying
arrival rate. Consequently, the estimators designed for spatial point
processes are not always applicable to biophysical data. For example, the
asymptotic variance formulas derived in the spatial context do not work for
high intensity photon arrival data. The general cross-validation method for
bandwidth selection [\citet{Guan2007}], due to its intense computation, does not
work well either for large photon arrival data. Furthermore, because of the
large arrival rate, the statistical performance of the kernel estimate
depends not only on the Poisson variation of $N(t)$ given $\lambda (t)$ but,
more importantly, on the stochastic properties of $\lambda (t)$. For
instance, we shall see in Section \ref{sec4} that the kernel estimate of
ACF will have asymptotically normal distribution only if $\lambda (t)$ has
short-range dependence.

Similar to classical kernel estimation, there is a bandwidth selection
problem associated with kernel inference of Cox process data. Using the mean
integrated square error (MISE) criterion [\citet{MarTsy1995}; \citet{Jones1996};
 \citet{Grund1994}; \citet{MarWand1992}; \citet{ParTur1992}; \citet{Diggle2003}], we propose a stable and fast
regression plug-in method to choose the bandwidth.

As our study is motivated by the analysis of scientific data, we apply our
method to photon arrival data from real biophysical experiments. The result
from our nonparametric inference helps elucidate the \textit{stochastic}
dynamics of proteins. In particular, our results show that as proteins
(such
as enzymes) spontaneously change their three-dimensional conformation, the
conformational fluctuation covers a very broad range of time scales,
highlighting the complexity of proteins' conformational dynamics.

The rest of the paper is organized as follows. Section \ref{sec2} considers kernel
estimation of the arrival rate $\lambda (t)$. Section \ref{sec3} focuses on
estimating the ACF of $\lambda (t)$, and provides some guidelines for
practical estimation. Section \ref{sec4} investigates the asymptotic distribution of
our kernel estimates, laying down the results for confidence interval
construction. In Section \ref{Numerical} we apply our method to simulated data and photon
arrival data from two biophysical experiments. We conclude in Section~\ref{sec6} with
some discussion and future work. The technical proofs are provided in the
supplementary material [\citet{ZhaKou2010}].

\section{Kernel estimation of the arrival rate}\label{sec2}

\subsection{The estimator}

Suppose within a time window $[0,T]$ a sequence of arrival times $
s_{1},s_{2},\ldots ,s_{K}$ has been observed from a Cox process $N(t)$,
which has stochastic arrival rates $\lambda (t)$. The goal is to infer from
the arrival times the stochastic properties of $\lambda (t)$. To do so, we
assume the following:

\begin{assumption}\label{ass1}
The arrival rate $\lambda (t)$ is a stationary and
ergodic process with finite fourth moments.
\end{assumption}

Stationarity (i.e., the distribution of $\{\lambda (t),t\in \mathsf{R}\}$ is
time-shift invariant) and ergodicity (i.e., essentially $\frac{1}{T}\int_{0}^{T}\lambda (s)\,ds\rightarrow E[\lambda (0)]$, as $T\rightarrow
\infty $) are both natural and necessary for making nonparametric inference
of $\lambda (t)$ from a single sequence of arrival data. Assumption \ref{ass1} is
particularly relevant for single-molecule biophysical experiments [\citet{Kou2009}] in which the system under study is typically in equilibrium or steady
state.

With Assumption \ref{ass1}, we now construct a kernel based arrival rate estimator
\begin{equation} \label{lambda}
\hat{\lambda}_{h}(t)=\sum_{i=1}^{K}f_{h}(t,s_{i}),\qquad \mbox{with } f_{h}(t,s)=\frac{1}{h}f\biggl(\frac{1}{h}(s-t)\biggr),
\end{equation}
where $f$ is a symmetric density function, and $h$ is the bandwidth. When $f$
is taken to be the uniform kernel, $\hat{\lambda}_{h}(t)$ amounts to the
binning-counting method used in the biophysics literature [Yang and Xie
(\citeyear{YanXie2002a}, \citeyear{YanXie2002b})], in which $\lambda (t)$ is estimated by the number of data
points falling into the bin containing $t$ divided by the bin width [see
also \citet{Diggle1985}; \citet{BerDig1989}]. One undesirable consequence of
uniform kernel is that, as points move in and out of the bins, $\hat{\lambda}_{h}(t)$ is artificially discontinuous. We thus consider general $f$, and
without loss of generality, we assume the following:

\begin{assumption}\label{ass2}
$f$ is a density function symmetric around 0 with
bounded support $[-b,b]$.
\end{assumption}

The assumption of bounded support in fact can be relaxed---essentially all
the results in this paper can be extended to kernels with unbounded
supports. However, to make the theory more presentable and to reduce the
length of algebra, we will work with Assumption \ref{ass2}.

When $t$ is getting too close to the boundaries of the observational time
window $[0,T]$, there are apparently not enough data to estimate $\lambda
(t) $ accurately. One method is to use end correction [see \citet{Diggle1985};
\citet{BerDig1989}]: $\hat{\lambda}^{B}(t)=\sum_{i=1}^{k}f_{h}(t,s_{i})/
\int_{0}^{T}f_{h}(t,s)\,ds$, which is identical to (\ref{lambda}) if $t\in[bh,T-bh]$.
However, the variance of the end-corrected-estimate $\hat{
\lambda}^{B}(t)$ tends to be large when $t$ is close to $0$ or $T$. We,
instead, estimate $\lambda (t)$ by
\begin{eqnarray} \label{lambda2}
\hat{\lambda}_{h}(t)=
\cases{
\displaystyle{\sum_{i=1}^{K}f_{h}(t,s_{i})},&\quad if $bh\leq t\leq T-bh$, \cr
\hat{\lambda}_{h}(bh),&\quad if $0 \leq t<bh$, \cr
\hat{\lambda}_{h}(T-bh),&\quad if $T-bh<t\leq T$,
}
\end{eqnarray}
that is, we use $\hat{\lambda}_{h}(bh)$ and $\hat{\lambda}_{h}(T-bh)$ to
approximate $\lambda (t)$ near the boundaries. We shall see shortly (Table \ref{tab1}) that for typical biophysical data the bandwidth $h$ is quite
small; thus, the bias of (\ref{lambda2}) is also small.

Since the choice of the bandwidth $h$ affects the performance of the kernel
estimate, we next determine the optimal $h$ that gives the smallest mean
integrated square error (MISE)
\[
\mathit{MISE}_{f}(h)=E\biggl( \frac{1}{T}\int_{0}^{T}\bigl(\hat{\lambda}_{h}(t)-\lambda
(t)\bigr)^{2}\,dt\biggr) .
\]
Owing to the stationarity and ergodicity of $\lambda (t)$, we have
\[
\mathit{MISE}_{f}(h)=E\bigl(\hat{\lambda}_{h}(t_{0})-\lambda (t_{0})\bigr)^{2}+O(h/T),
\]
where $t_{0}$ is any number within $[bh,T-bh]$, say, $t_{0}=T/2$, and the $
O(h/T)$ term arises from the boundary of $[0,T]$. Hence, minimizing the MISE
amounts to minimizing the MSE of $\hat{\lambda}_{h}(t_{0})$.

Let $C(t)$ denote the ACF of the arrival rate $\lambda (t)$: $C(t)=\operatorname{cov}(\lambda (0),\lambda (t))$.
To find the optimal bandwidth that minimizes
the MSE of $\hat{\lambda}_{h}(t_{0})$, we make one more assumption.
\begin{assumption}\label{ass3}
The ACF $C(t)$ is twice continuously differentiable
for $t>0$, and has nonzero right derivative at 0, that is, $C^{\prime
}(0^{+})=\lim_{s\rightarrow 0^{+}}(C(s)-C(0))/s$ exits and is nonzero.
\end{assumption}

This assumption reflects the fact that the arrival rate process $\lambda (t)$
in real experiments is usually not differentiable [\citet{Parzen1962}, Chapter 3];
for example, $\lambda (t)$ could be a finite-state continuous Markov chain,
whose path consists of piecewise jumps and whose ACF is a mixture of
exponential functions, which are nondifferentiable at zero, or $\lambda (t)$
could be a functional of a stationary nondifferentiable Gaussian process,
such as the Ornstein--Uhlenbeck process (representing a harmonic oscillator).
\begin{theorem}\label{theorem1}
Under Assumptions \ref{ass1}--\ref{ass3}, the MSE of $\hat{\lambda}_{h}(t_{0})$
is given by
\begin{equation}\label{MSE}
\qquad E\bigl(\hat{\lambda}_{h}(t_{0})-\lambda (t_{0})\bigr)^{2}=\frac{1}{h}E(\lambda
(0))\int_{-b}^{b}f^{2}(r)\,dr+hC^{\prime }(0^{+})\gamma _{f}+R_{2}(h),
\end{equation}
where the constant
\begin{equation} \label{rf}
\gamma_{f}=\int_{-b}^{b}\int_{-b}^{b}f(r_{1})f(r_{2})|r_{1}-r_{2}|\,dr_{1}\,dr_{2}-2
\int_{-b}^{b}f(r)|r|\,dr<0,
\end{equation}
and
\begin{eqnarray*}
R_{2}(h)&=&\int_{-b}^{b}\int_{-b}^{b}f(r_{1})f(r_{2})
\int_{0}^{|r_{1}-r_{2}|h}(|r_{1}-r_{2}|h-s)C^{\prime \prime
}(s)\,ds\,dr_{1}\,dr_{2} \\
&&{}-2\int_{-b}^{b}f(r)\int_{0}^{|r|h}(|r|h-s)C^{\prime \prime }(s)\,ds\,dr =
o(h).
\end{eqnarray*}
The optimal $h$ that minimizes the sum of the first two terms (i.e., the
main terms) of the right-hand side of (\ref{MSE}) is given by
\begin{equation}\label{hf}
h_{\mathrm{opt}}=\biggl[\frac{E(\lambda (0))}{C^{\prime }(0^{+})\gamma _{f}}
\int_{-b}^{b}f^{2}(r)\,dr\biggr]^{1/2}.
\end{equation}
\end{theorem}

The constant $\gamma _{f}$ is strictly negative as long as $f$ is a density
function. $R_{2}(h)$ is the remainder term. For data with large arrival
rates, $h_{\mathrm{opt}}$ is small, and $R_{2}(h_{\mathrm{opt}})$ contributes little to the
MSE. The proof of the theorem is given in the supplementary material [\citet{ZhaKou2010}].

Since $h_{\mathrm{opt}}$ involves unknown quantities, for real applications we use a
regression based plug-in method to estimate it. First, $\mu =E(\lambda (0))$
is unbiasedly estimated by $\hat{\mu}=K/T$, the total number of arrivals
divided by the time window length, because
\[
E(\hat{\mu})=\frac{1}{T}E\{E[k|\lambda (\cdot)]\}=\frac{1}{T}
E\biggl\{\int_{0}^{T}\lambda (t)\,dt\biggr\}=E(\lambda (0)).
\]
Next, we use a regression method to estimate $C^{\prime }(0^{+})$. We will
discuss this regression estimate in detail in Section \ref{Sec3.2} when we
study the ACF estimation. Plugging $\hat{\mu}$ and $\hat{C}^{\prime }(0^{+})$
into (\ref{hf}) yields our estimate $\hat{h}_{\mathrm{opt}}$.

\subsection{Numerical illustration}

We use two simulation examples to illustrate our method. In the first
example, the arrival rate $\lambda (t)$ of the Cox process follows a
continuous-time two-state Markov chain, which can be depicted as
\begin{equation}\label{2state}
A\mathop{\rightleftarrows}_{k_2}^{k_1} B,
\end{equation}
where $k_{1}$ and $k_{2}$ represent the transition rates between the two
states $A$ and $B$. This model has been used in the chemistry and biophysics
literature [\citet{ReiSki1994}] to model spectral and fluorescence data
from two-level systems, such as the open-close of a DNA hairpin [\citet{KouXie2005a}],
 and the on-off of ion channels [\citet{Hawkes2005}; Sakmann and Neher (\citeyear{SakNeh1995})]. We set the
  transition rates $k_{1}=2$, $k_{2}=5$ and arrival
rates $\lambda _{A}=1000$ and $\lambda _{B}=400$ respectively at states $A$
and $B$ in the simulation; the observational time $T=500$. These numbers are
taken to mimic a typical photon arrival experiment in biophysics. We
generate a realization of $\lambda (t)$ from the two-state model and then
the arrival times $s_{i}$, $i=1,\ldots ,K$ on top of it. The true mean
arrival rate $\mu =E(\lambda (0))=(k_{2}\lambda _{A}+k_{1}\lambda_{B})/(k_{1}+k_{2})$
is equal to $828.57$ in this case. The simulated data
has the empirical mean arrival rate $\hat{\mu}=K/T=823.11$.

\begin{figure}

\includegraphics{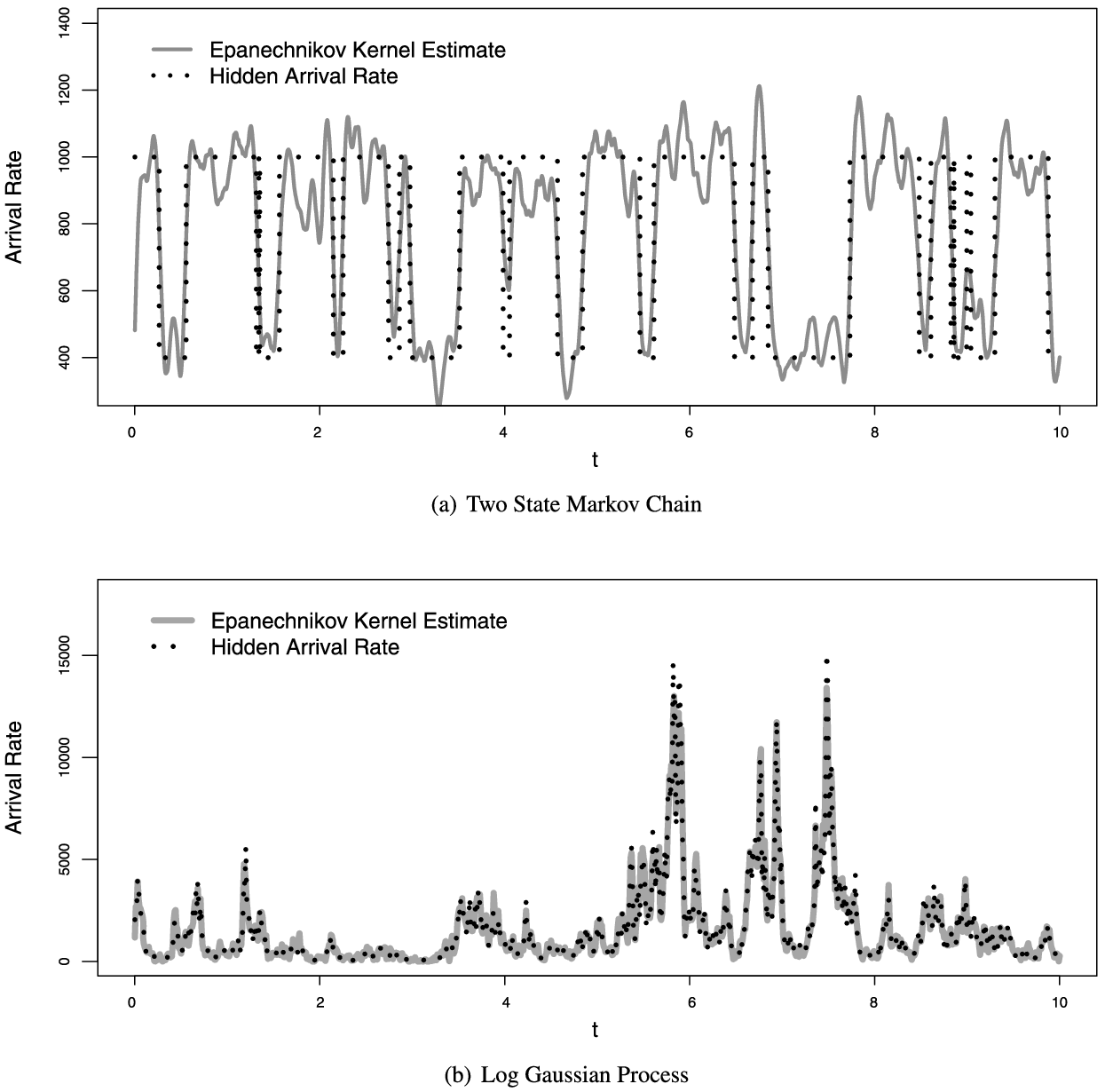}

\caption{Arrival rate estimation. \textup{(a)} Two state Markov chain
with $\mu =828.57$ and $T=500$. \textup{(b)} Log Gaussian $\lambda (t)$ with
$\gamma (t)= 1/(1+|t|)^6$ and $T=1500$.}\label{fig1}
\end{figure}

Figure \ref{fig1}(a) shows the estimate $\hat{\lambda}_{\hat{h}
_{\mathrm{opt}}}(t)$, compared with the true $\lambda (t)$, based on the Epanechnikov
kernel $f(t)=\frac{3}{4}(1-t^{2})I(|t|\leq 1)$.
Figure \ref{fig1}(a) represents a typical result. We see that $\lambda (t)$
is well recovered. Table \ref{tab1} summarizes the results based on 100
independent simulations for applying our method with four different kernels:
the uniform, Epanechnikov, triangular $f(t)=(1-|t|)I(|t|\leq 1)$ and quartic
$f(t)=\frac{15}{16}(1-t^{2})^{2}I(|t|\leq 1)$ kernels. The second and third
columns present the optimal bandwidth $h_{\mathrm{opt}}$ for each kernel and their
estimates $\hat{h}_{\mathrm{opt}}$ obtained through the regression plug-in procedure.
The next four columns show the normalized empirical MISE $\frac{1}{T\hat{\mu}
^{2}}\int_{0}^{T}(\hat{\lambda}_{h}(t)-\lambda (t))^{2}\,dt$ for $h=h_{\mathrm{opt}}$, $
\hat{h}_{\mathrm{opt}}$, $h_{\mathrm{opt}}/2$ and $2h_{\mathrm{opt}}$ respectively. It is noticeable
that (i) the regression plug-in method for approximating $h_{\mathrm{opt}}$ works
well; in particular, the empirical MISE with the estimated $\hat{h}_{\mathrm{opt}}$
is close to that of $h_{\mathrm{opt}}$; (ii) the performance of the kernel method
largely depends on the choice of the bandwidth; if one uses a nonoptimal
bandwidth, such as twice or half $h_{\mathrm{opt}}$, the error can increase as much
as $30\%$; (iii) the choice of the kernel is not as crucial as that of the
bandwidth, which echoes the classical result in kernel density estimation;
and (iv) the widely used binning method, which is equivalent to using the
uniform kernel, gives the largest error.

\begin{table}
\tabcolsep=0 pt
\caption{Kernel estimates of the arrival rate for the two-state
continuous-time Markov chain model. The second and third columns present the
optimal bandwidth $h_{\mathrm{opt}}$ and the mean of their estimates $\hat{h}_{\mathrm{opt}}$
based on 100 simulations. The next several columns show the mean of the
normalized empirical MISE $\frac{1}{T\hat{\mu}^{2}}\int
_{0}^{T}(\hat{\lambda}_{h}(t)-\lambda (t))^{2}\ dt$ for $
h=h_{\mathrm{opt}}$, $\hat{h}_{\mathrm{opt}}$, $h_{\mathrm{opt}}/2$ and $2h_{\mathrm{opt}}$
respectively. The
numbers in the brackets are the associated standard deviations}\label{tab1}
\begin{tabular*}{\textwidth}{@{\extracolsep{\fill}}lcccccc@{}}
\hline
& \multicolumn{2}{c}{\textbf{Bandwidth (in $\bolds{10^{-2}}$)}} & \multicolumn{4}{c@{}}{\textbf{Normalized empirical MISE (in $\bolds{10^{-2}}$)}}\\[- 7pt]
&\multicolumn{2}{c}{\hrulefill}&\multicolumn{4}{c@{}}{\hrulefill}\\
\textbf{Kernel} $\bolds{f}$ & $\bolds{h_{\mathrm{opt}}}$ & $\bolds{\hat{h}_{\mathrm{opt}}}$ & $\bolds{h_{\mathrm{opt}}}$ & $\bolds{\hat{h}_{\mathrm{opt}}}$ & $
\bolds{h_{\mathrm{opt}}/2}$ & $\bolds{2h_{\mathrm{opt}}}$ \\
\hline
Uniform & 4.93 & 5.19 (0.34) & 2.43 (0.057) & 2.43 (0.059) & 3.05 (0.050) &
2.93 (0.092) \\
Epanechnikov & 6.40 & 6.67 (0.50) & 2.23 (0.053) & 2.24 (0.055) & 2.82
(0.048) & 2.70 (0.085) \\
Quartic & 7.74 & 8.07 (0.66) & 2.20 (0.052) & 2.21 (0.055) & 2.78 (0.047) &
2.65 (0.083) \\
Triangular & 7.33 & 7.64 (0.63) & 2.17 (0.052) & 2.17 (0.054) & 2.74 (0.047)
& 2.61 (0.082)\\
\hline
\end{tabular*}
\end{table}

In the second example, the arrival rate $\lambda (t)$ follows a log Gaussian
process: $\lambda (t)=M\exp (W(t))$, where $M>0$ and $W(t)$ is a stationary
zero-mean Gaussian process with the autocovariance function $\gamma (t)$. It
is straightforward to obtain $\mu =E(\lambda (t))=Me^{\gamma (0)/2}$ and $
C(t)=M^{2}e^{\gamma (0)}(e^{\gamma (t)}-1)$. We take $\gamma
(t)=1/(1+a|t|)^{H}$ so that $C(t)$ decreases at the order $t^{-H}$. Both $a$
and $H$ are positive constants. The log Gaussian process has been used to
model the conformational dynamics and reactivity of enzyme molecules [\citet{Min2005a}; \citet{KouXie2004}].
For instance, \citet{KouXie2004}
showed that an enzyme's conformational fluctuation can be modeled by a
generalized Langevin equation, in which the $\lambda (t)$ follows a log
Gaussian process with the ACF having a power law decay $C(t)\sim t^{-H}$.
Here, we take $H=6$, $a=1$, $M=1000$ and $T=1500$ in our simulation to mimic
a real photon arrival data of this kind. Figure \ref{fig1}(b)
compares the estimate $\hat{\lambda}_{\hat{h}_{\mathrm{opt}}}(t)$ to the true $\lambda (t)$ for the Epanechnikov kernel. We repeat the simulation 100
times. Figure~\ref{fig1}(b) represents a typical outcome. We
see that $\lambda (t)$ is well recovered. The detailed estimation results
are summarized in Table \ref{tab2}. Again, we can see that the regression
plug-in method for estimating $h_{\mathrm{opt}}$ works well and that the performance
of the kernel method depends largely on the choice of the bandwidth and less
so on the kernels. The widely used binning method again gives the poorest
result.

\begin{table}
\tabcolsep=0 pt
\caption{Kernel estimates of the arrival rate for the log Gaussian model
with $C(t)=1000^{2}\break\times(\exp (1/(1+|t|)^{6}+1)-e)$. The second and third columns
present $h_{\mathrm{opt}}$ and the mean of the estimate $\hat{h}_{\mathrm{opt}}$ based on 100
simulations. The next four columns show the mean of the normalized empirical
MISE for $h=h_{\mathrm{opt}}$, $\hat{h}_{\mathrm{opt}}$, $h_{\mathrm{opt}}/2$ and
$2h_{\mathrm{opt}}$
respectively. The numbers in the brackets are the associated standard
deviations}\label{tab2}
\begin{tabular*}{\textwidth}{@{\extracolsep{\fill}}ld{2.5}d{2.8}d{1.8}d{1.8}d{2.8}d{2.8}@{}}
\hline
& \multicolumn{2}{c}{\textbf{Bandwidth (in $\bolds{10^{-3}}$)}}
& \multicolumn{4}{c@{}}{\textbf{Normalized empirical MISE (in $\bolds{10^{-2}}$)}}\\[- 7 pt]
& \multicolumn{2}{c}{\hrulefill}&\multicolumn{4}{c@{}}{\hrulefill}\\
\textbf{Kernel} $\bolds{f}$ &\multicolumn{1}{c}{$\bolds{h_{\mathrm{opt}}}$} &\multicolumn{1}{c}{$\bolds{\hat{h}_{\mathrm{opt}}}$}& \multicolumn{1}{c}{$\bolds{h_{\mathrm{opt}}}$} &
\multicolumn{1}{c}{$\bolds{\hat{h}_{\mathrm{opt}}}$} &\multicolumn{1}{c}{$\bolds{h_{\mathrm{opt}}/2}$} & \multicolumn{1}{c@{}}{$\bolds{2h_{\mathrm{opt}}}$} \\
\hline
Uniform & 7.49 & 7.68\ (0.27) & 8.08\ (0.16) & 8.20\ (0.19) & 10.3\ (0.17) &
10.1\ (0.32) \\
Epanechnikov & 9.73 & 10.1\ (0.28) & 7.45\ (0.15) & 7.45\ (0.15) & 9.33\ (0.15)
& 9.26\ (0.30) \\
Quartic & 11.8 & 12.1\ (0.32) & 7.34\ (0.15) & 7.34\ (0.15) & 9.17\ (0.15) &
9.14\ (0.29) \\
Triangular & 11.1 & 11.4\ (0.30) & 7.23\ (0.15) & 7.23\ (0.15) & 9.12\ (0.15) &
8.97\ (0.29)\\
\hline
\end{tabular*}
\end{table}

\section{Estimating the ACF}\label{sec3}

\subsection{Kernel estimation}

In this section we consider kernel estimation of the ACF $C(t)$ of the
arrival rate. The ACF is useful in exploring the dependence structure of new
stochastic dynamics and identifying appropriate parametric models for the
data.

For example, most ion channel dynamics and most chemical reactions involve
reversible transitions among the various discrete chemical states in which
the system can exist. In these systems, a fast decay of the ACF, such as an
exponential decay, indicates that the transition among the discrete states
has a short memory, and the underlying biological process has a relatively
simple mechanism, such as having only two or three states. In the case of
ion channels, the simplest dynamic consists of a transition between a single
shut state of the ion channel and a single open state [\citet{SakNeh1995},
\citet{Hawkes2005}], in which the ACF is a single exponential function over
time. In the case of a protein's conformational fluctuation, the simplest
scenario is a transition between two distinct conformation states (where the
protein reversibly and spontaneously crosses the energy barrier that
separates the two states). In the case of enzyme catalytic fluctuations, the
simplest scenario is that the enzyme interconverts among a small numbers of
states, in which the ACF has a near exponential decay [\citet{Schenter1999};
Yang and Xie (\citeyear{YanXie2002a}, \citeyear{YanXie2002b});
 \citet{KouXie2005a}; \citet{Kou2005b}].

A slow decay of ACF, such as a power-law decay, on the other hand, signifies
a complicated process and points to an intricate internal structure, such as
the existence of a large number of conformation states or the presence of a
complicated energy landscape [\citet{KouXie2004}; \citet{Min2005b}].

To ease the presentation, we first consider the situation where the mean
arrival rate $\mu =E(\lambda (0))$ is known, and later relax the results for
unknown $\mu $. The basic idea is as follows. If we actually observe the
realization of $\lambda (t)$, then using its ergodicity property, we have a
natural estimate $\frac{1}{T-t}\int_{0}^{T-t}(\lambda (s)-\mu)(\lambda
(s+t)-\mu)\,ds$ for $C(t)$. Now $\lambda (s)$ is unobserved; we replace it by
$\hat{\lambda}_{h}(s)$. To avoid the bias at the boundary of the observation
window, our kernel estimate of $C(t)$ is
\begin{eqnarray}\label{C(t)}
\hat{C}_{\mu ,h}(t)=\frac{1}{T-2bh-t}\int_{bh}^{T-bh-t}\bigl(\hat{\lambda}_{h}(s+t)-\mu \bigr)\bigl(\hat{\lambda}_{h}(s)-\mu
\bigr)\,ds,\nonumber\\ [-8 pt] \\ [-8 pt]
\eqntext{t\in [0,T-2bh).}
\end{eqnarray}
The next two lemmas tell us the bias and variance of $\hat{C}_{\mu ,h}(t)$
for estimating $C(t)$ at a fixed $t$.

\begin{lemma}
\label{LemmaBias}Under Assumptions \ref{ass1} and \ref{ass2},
\begin{eqnarray} \label{bias}
E(\hat{C}_{\mu ,h}(t))&=&\frac{\mu}{h}\int_{-b}^{b}f\biggl(r+\frac{t}{h}\biggr)f(r)\,dr\nonumber\\ [-8pt]\\ [- 8 pt]
&&{}+\int_{-b}^{b}\int_{-b}^{b}C\bigl(|t+(r-m)h|\bigr)f(r)f(m)\,dr\,dm.\nonumber
\end{eqnarray}
Furthermore, if Assumption \ref{ass3} also holds, then
\begin{eqnarray*}
&&E(\hat{C}_{\mu ,h}(t))\\
&&\qquad=\cases{\displaystyle
C(t)+C^{\prime \prime }(t)h^{2}\int_{-b}^{b}r^{2}f(r)\,dr+o(h^{2}), & \quad $t\geq 2bh$;\cr
\displaystyle{C(0)+\frac{\mu}{h}\int_{-b}^{b}f\biggl(r+\frac{t}{h}\biggr)f(r)\,dr}\cr
\quad\displaystyle{}+C^{\prime}(0^{+})\int_{-b}^{b}\int_{-b}^{b}|t+(r-m)h|f(r)f(m)\,dr\,dm\cr
\quad{}+o(h), & \quad $t\in[0,2bh)$.}
\end{eqnarray*}
\end{lemma}

The bias of the ACF estimate $\hat{C}_{\mu ,h}(t)$ is due to the fact that $
\lambda (t)$ is estimated by ``borrowing''
information from the neighboring regions. When $t<2bh$, the data points used
to calculate $\hat{\lambda}_{h}(s)$ and $\hat{\lambda}_{h}(s+t)$ overlap,
resulting in the extra bias $\mu \int_{-b}^{b}f(r+t/h)f(r)\,dr/h.$

For notational convenience, we denote
\begin{eqnarray*}
C_{3}(t_{1},t_{2}) &=&E\bigl\{\bigl(\lambda (0)-\mu \bigr)\bigl(\lambda (t_{1})-\mu \bigr)\bigl(\lambda
(t_{2})-\mu \bigr)\bigr\},\\
v(t_{1},t_{2},s^{\prime }-s) &=&\operatorname{cov}\bigl\{\bigl(\lambda (s)-\mu \bigr)\bigl(\lambda
(s+t_{1})-\mu \bigr),\bigl(\lambda (s^{\prime })-\mu \bigr)\bigl(\lambda (s^{\prime }+t_{2})-\mu
\bigr)\bigr\}.
\end{eqnarray*}
Because of the stationarity of $\lambda (t)$, $C_{3}(t_{1},t_{2})=C_{3}(t_{2},t_{1})$ and $v(t_{1},t_{2},s^{\prime
}-s)=v(t_{2},t_{1},s-s^{\prime })$. The following two technical assumptions
are needed to characterize the asymptotic behavior of $\operatorname{var}(\hat{C}_{\mu ,h}(t))$.

\begin{assumption}\label{ass4}
The three-step correlation $C_{3}(t_{1},t_{2})$ is
continuous and satisfies
\[
\lim_{|t_{2}|\rightarrow \infty }C_{3}(t_{1},t_{2})=0\qquad\mbox{for any fixed }
t_{1}.
\]
\end{assumption}
\begin{assumption} \label{ass5}
The cross correlation $v(t_{1},t_{2},s)$ is
continuous and satisfies
\[
\lim_{|s|\rightarrow \infty }v(t_{1},t_{2},s)=0\qquad\mbox{for any fixed }t_{1}\mbox{ and }t_{2}.
\]
\end{assumption}

These two assumptions reflect the intuitive idea that as time-points move
far away from each other, their dependence should eventually vanish. They
are satisfied by most stationary and ergodic processes that one encounters
in practice, such as continuous-time finite-state Markov Chains and
functionals of stationary and ergodic Gaussian processes.

\begin{lemma}\label{LemmaVar}
The variance of $\hat{C}_{\mu ,h}(t)$ can be decomposed as
\begin{equation}\label{vardec}
\operatorname{var}(\hat{C}_{\mu ,h}(t))=\operatorname{var}\{E(\hat{C}_{\mu,h}(t)|\lambda
(\cdot))\}+E\{\operatorname{var}(\hat{C}_{\mu ,h}(t)|\lambda
(\cdot))\}.
\end{equation}
Under Assumptions 1, 2, 4 and 5,
\begin{eqnarray}\label{var1}
\qquad&&\operatorname{var}\{E(\hat{C}_{\mu ,h}(t)|\lambda(\cdot))\}\nonumber\\[- 8pt]\\[- 8pt]
&&\qquad=\frac{1}{(T-2bh-t)^{2}}\biggl(A_{t,T}^{h}+\frac{2}{h}B_{t,T}^{h}+\frac{1}{h^{2}}
C_{t,T}^{h}\biggr)\rightarrow 0,\qquad\mbox{as }T\rightarrow \infty,\nonumber
\end{eqnarray}
where
\begin{eqnarray*}
&&A_{t,T}^{h}
=\int\!\!\!\int_{[bh,T-bh-t]^{2}}\biggl[\int\!\!\!\int\!\!\!\int\!\!\!\int_{[-b,b]^{4}}v\bigl(t+(l-m)h,t+(l^{\prime}-m^{\prime })h,\\
&&\hspace{63pt}\hphantom{A_{t,T}^{h}=\int\!\!\!\int_{[bh,T-bh-t]^{2}}\biggl[\int\!\!\!\int\!\!\!\int\!\!\!\int_{[-b,b]^{4}}}s^{\prime }-s+(l^{\prime }-l)h\bigr)f(l)\\
&&\hspace{6pt}\hphantom{A_{t,T}^{h}=\int\!\!\!\int_{[bh,T-bh-t]^{2}}\biggl[\int\!\!\!\int\!\!\!\int\!\!\!\int_{[-b,b]^{4}}}{}\times f(m) f(l^{\prime })f(m^{\prime })\ dl\ dm\ dl^{\prime }\ dm^{\prime
}\biggr]\ ds\ ds^{\prime }, \\
&&B_{t,T}^{h}=\int\!\!\!\int_{[bh,T-bh-t]^{2}}\biggl[\int\!\!\!\int\!\!\!\int\!\!\!\int_{[-b,b]^{3}}C_{3}\bigl(t+(l^{\prime
}-m^{\prime })h,\\
&&\hphantom{B_{t,T}^{h}=\int\!\!\!\int_{[bh,T-bh-t]^{2}}\biggl[\int\!\!\!\int\!\!\!\int\!\!\!\int_{[-b,b]^{3}}C_{3}\bigl(}s-s^{\prime }+t+(l-m^{\prime })h\bigr)f(l^{\prime }) \\
&&\hspace{-20pt}\hphantom{B_{t,T}^{h}=\int\!\!\!\int_{[bh,T-bh-t]^{2}}\biggl[\int\!\!\!\int\!\!\!\int\!\!\!\int_{[-b,b]^{3}}C_{3}\bigl(}\times f(m^{\prime })f(l)f\biggl(l+\frac{t}{h}\biggr)\ dl\ dm^{\prime}\ dl^{\prime }\biggr]\ ds\ ds^{\prime }, \\
&&C_{t,T}^{h} =\int\!\!\!\int_{[bh,T-bh-t]^{2}}\biggl[\int\!\!\!\int_{[-b,b]^{2}}C\bigl(s^{\prime}-s+(l^{\prime }-l)h\bigr)f(l)\\
&&\hphantom{C_{t,T}^{h}=\int\!\!\!\int_{[bh,T-bh-t]^{2}}\biggl[\int\!\!\!\int_{[-b,b]^{2}}}{}\times f\biggl(l+\frac{t}{h}\biggr)f(l^{\prime })f\biggl(l^{\prime }+\frac{t}{h}\biggr)\ dl\ dl^{\prime }
\biggr]\ ds\ ds^{\prime},
\end{eqnarray*}
and
\begin{equation}\label{var2}
\qquad E\{\operatorname{var}(\hat{C}_{\mu ,h}(t)|\lambda (\cdot))\}=\frac{1}{T-2bh-t}
\biggl(D_{t}^{h}+\frac{1}{h}E_{t}^{h}+\frac{1}{h^{2}}F_{t}^{h}\biggr)+O\biggl(\frac{h}{T^{2}}\biggr),
\end{equation}
where the three terms $D_{t}^{h},E_{t}^{h}$ and $F_{t}^{h}$ do not depend on
$T$. Their exact but lengthy expressions, involving multiple integrals, are
given in the supplementary material [Zhang and Kou (\citeyear{ZhaKou2010})].
\end{lemma}

Equation (\ref{vardec}) indicates that the variance of the ACF estimate
arises from two sources: the Poisson variation---$E\{\operatorname{var}(\hat{C}
_{\mu ,h}(t)|\lambda (\cdot))\}$---and the variation from $\lambda (t)$ --
$\mathrm{var}\{E(\hat{C}_{\mu ,h}(t)|\lambda (\cdot))\}$. $A_{t,T}^h$ is
the main part of $\mathrm{var}\{E(\hat{C}_{\mu ,h}(t)|\lambda (\cdot))\}$.
When $t>2bh$, $B_{t,T}^h$ and $C_{t,T}^h$ both equal zero.

Based on Lemmas \ref{LemmaBias} and \ref{LemmaVar}, the next theorem tells
us that as the observation time $T$ gets larger, $\hat{C}_{\mu ,h}(t)$
consistently estimates $C(t)$.

\begin{theorem}
\label{theorem2}Suppose that $C(t)$ is a continuous function of $
t\in[0,\infty)$ and that Assumptions \ref{ass1}, \ref{ass2}, \ref{ass4} and \ref{ass5} hold. Then for any fixed
$t>0$, as $T\cdot h\rightarrow \infty $ and $h\rightarrow 0$,
\[
\hat{C}_{\mu ,h}(t)\rightarrow C(t)\qquad\mbox{in }L^{2},
\]
so, in particular,
\[
\hat{C}_{\mu ,h}(t)\rightarrow C(t)\qquad\mbox{in probability}.
\]
\end{theorem}

The assumption of continuous $C(t)$ is satisfied for general continuous-time
stationary and ergodic processes. We note that in the context of spatial
point processes, different estimates of the covariance function have been
proposed [see, e.g., \citet{StoSto1994} and \citet{Diggle2003}]. We use (\ref{C(t)})
here mainly due to its internal coherency: an estimate of $\lambda (t)$ naturally leads to an estimate of $C(t)$.

\subsection{Practical consideration}\label{Sec3.2}

To use the kernel estimate in practice, a few issues arise naturally.

\subsubsection*{Unknown $\mu $} In real applications, the mean arrival
rate $\mu $ is unknown. Employing its unbiased estimate $\hat{\mu}=K/T$, we
use
\[
\hat{C}_{\hat{\mu},h}(t)=\frac{1}{T-2bh-t}\int_{bh}^{T-bh-t}\bigl(\hat{\lambda}
_{h}(s+t)-\hat{\mu}\bigr)\bigl(\hat{\lambda}_{h}(s)-\hat{\mu}\bigr)\,ds
\]
to estimate the ACF. A question follows immediately: is $\hat{C}_{\hat{\mu}
,h}(t)$ still a consistent estimator? The next theorem provides a positive
answer.

\begin{theorem}\label{theorem2.1}
Suppose that $C(t)$ is a continuous function of $
t\in[0,\infty)$ and that Assumptions \ref{ass1}, \ref{ass2}, \ref{ass4} and \ref{ass5} hold. Then for any fixed
$t>0$, as $T\cdot h\rightarrow \infty $ and $h\rightarrow 0$,
\[
\hat{C}_{\hat{\mu},h}(t)\rightarrow C(t)\qquad\mbox{in probability}.
\]
\end{theorem}

\subsubsection*{Bias correction for small $t$} From Lemma \ref{LemmaBias}, we see
that $\hat{C}_{\mu ,h}(t)$ has an extra bias $\mu
\int_{-b}^{b}f(r+t/h)f(r)\,dr/h$ for $t<2bh$. A bias correction can be
conducted for $t<2bh$, yielding
\begin{equation} \label{BiasCorrectCt}
\tilde{C}_{\hat{\mu},h}(t)=\hat{C}_{\hat{\mu},h}(t)-\frac{\hat{\mu}}{h}
\int_{-b}^{b}f\biggl(r+\frac{t}{h}\biggr)f(r)\,dr.
\end{equation}

\subsubsection*{Estimating $h_{\mathrm{opt}}$} 
\begin{figure}[b]

\includegraphics{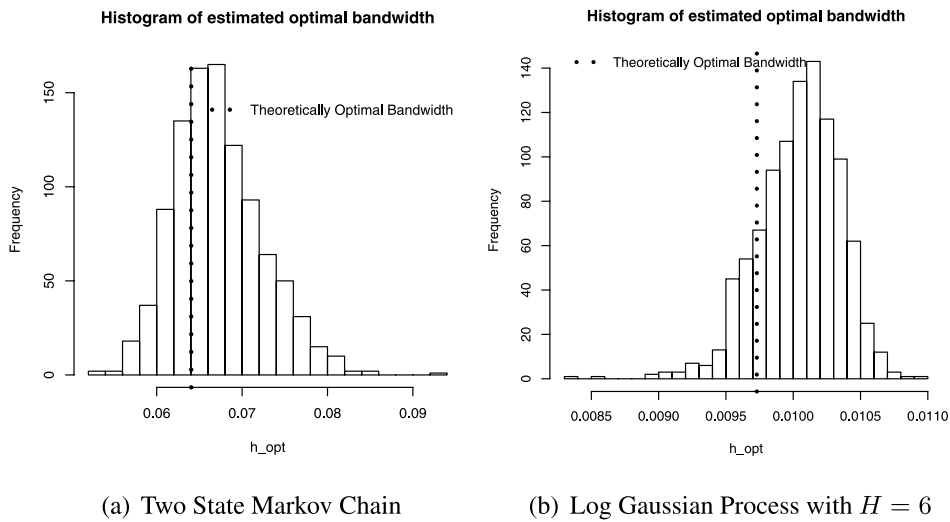}

\caption{Estimating the optimal bandwidth with the Epanechnikov kernel.}\label{hEpan}
\end{figure}

In Section \ref{sec2} we briefly
described how to estimate $h_{\mathrm{opt}}$ to recover the arrival rate, where the
key is to estimate the derivative $C^{\prime }(0^{+})$. With Lemma \ref
{LemmaBias} established, we now explain our estimate in detail. Lemma \ref
{LemmaBias} tells us that, for small $t$, the expectation of $\tilde{C}_{\hat{
\mu},h}(t)$ depends on $\int_{-b}^{b}\int_{-b}^{b}|t+(r-m)h|f(r)f(m)\,dr\,dm$
linearly with $C^{\prime }(0^{+})$ as the slope. This suggests that we can
calculate $Y_{i}=\tilde{C}_{\hat{\mu},h}(t_{i})$ for evenly spaced $t_{i}\in
\lbrack 0,2bh)$, say, ten points, and regress $Y_{i}$ on $X_{i}=\int_{-b}^{b}
\int_{-b}^{b}|t_{i}+(r-m)h|f(r)f(m)\,dr\,dm$. The regression slope is our
estimate of $C^{\prime }(0^{+})$. Compared to the naive idea of using a
numerical derivative $(\hat{C}(\Delta)-\hat{C}(0))/\Delta $ for some small $
\Delta $ to approximate $C^{\prime }(0^{+})$, this regression estimate is
not only easy to implement, but, more importantly, much stabler in
performance (see Figure \ref{hEpan}).

For the calculation of $\tilde{C}_{\hat{\mu},h}(t_{i})$, one needs to start
from an initial $h$. We use $h=\rho /\hat{\mu}=\rho T/K$, where the constant
$\rho $ (for example, between 3 and 10) is the average number of data points
falling in an interval of length $h$. This choice of initial $h$ ensures
that there are enough points in the kernel to give reliable $\tilde{C}_{\hat{
\mu},h}(t_{i})$. Throughout our simulation and real data analysis, where $
\hat{\mu}$ is in the hundreds, we found that taking $\rho $ between 3 and 10
gives almost identical results. Figure \ref{hEpan} shows how our estimate $
\hat{h}_{\mathrm{opt}}$ behaves for the two simulation examples of Section \ref{sec2}:
 the two-state Markov chain and the log Gaussian process model. The dotted
vertical line is the true $h_{\mathrm{opt}}$. The histograms in Figure \ref{hEpan}
are based on 1000 i.i.d. replications of the Cox process. The estimate $\hat{
h}_{\mathrm{opt}}$ is seen to be stable and close to $h_{\mathrm{opt}}$.

\subsubsection*{The bandwidth $h$ for estimating $C(t)$} To
estimate the ACF $C(t)$, a natural question is the choice of $h$. It could
be different from that associated with recovering the arrival rate. One
approach might be as follows: based on the results of Lemmas \ref{LemmaBias}
and \ref{LemmaVar}, find the asymptotic leading terms in bias-square and
variance, and then search for $h$ to optimize their linear combination.
Although conceptually ``simple'', this
approach in fact has several major difficulties that make it ineffective for
practical use: (a) As the equations in Lemmas \ref{LemmaBias} and \ref
{LemmaVar} involve fourth moment and covariance of $\lambda (t)$, one needs
to estimate them. Since the estimates of high moments often have large
variability, the resulting $h$ tends to be highly variable. (b) The bias and
variance formulas depend on the specific value of $t$, which implies that
for each $t$ there is an $h$. Consequently, if the entire curve $C(t)$ is of
interest (as in many scientific studies), the computation becomes very
intensive. (c) In order for the bias-squared to become comparable to the\
variance and in order for the asymptotics to take effect, $h$ needs to be
much smaller than $(C(t)+\mu ^{2})/(\int_{0}^{T}v(t,t,s)\,ds)$, which in turn
requires $T$ to be quite large: $T\sim O((C(t)+\mu ^{2})/(C^{\prime \prime
}(t)^{2}h^{5}))$. However, real biophysical data with large $\mu $ and
moderate $T$ hardly satisfy this requirement. (d) In order to identify the
asymptotic leading terms, more technical assumptions, such as short-range
dependence of $\lambda (t)$ [i.e., $\int_{0}^{\infty }C(t)\,dt<\infty $], have
to be imposed, which restricts the estimate's general applicability.

For these reasons, we recommend using $\hat{h}_{\mathrm{opt}}$ for $t\geq 2b\hat{h}
_{\mathrm{opt}}$ and a smaller bandwidth $h=\min (\rho /\hat{\mu},\hat{h}_{\mathrm{opt}})$,
where $\rho \in \lbrack 3,10]$ for $t<2b\hat{h}_{\mathrm{opt}}$ to estimate the ACF $
C(t)$. The reason to use $\min (\rho /\hat{\mu},\hat{h}_{\mathrm{opt}})$ instead of $
\hat{h}_{\mathrm{opt}}$ for $t<2b\hat{h}_{\mathrm{opt}}$ is that for large mean arrival rate $
\mu $, $\rho /\hat{\mu}$ can be smaller than $\hat{h}_{\mathrm{opt}}$; in this case
Lemma \ref{LemmaBias} tells us that for small $t$ the bias of $\tilde{C}_{
\hat{\mu},h}(t)$ from $h=\rho /\hat{\mu}$ tends to be smaller than that of $
\hat{h}_{\mathrm{opt}}$, while Lemma \ref{LemmaVar} indicates that the variances of
the two are about the same. Thus, for small $t$, $\min (\rho /\hat{\mu},\hat{
h}_{\mathrm{opt}})$ appears to be a better choice. Although our bandwidth
recommendation does not guarantee the smallest MSE for $C(t)$ at every $t$,
it does offer a stable and easy-to-compute bandwidth. We will demonstrate
the effectiveness of this choice in Section \ref{Numerical} (see Table \ref{tab3}) when we study confidence interval construction.

\subsubsection*{Approximating the variance of $\hat{C}_{\mu ,h}(t)$} For estimating
the variance of $\hat{C}_{\mu ,h}(t)$ (e.g., in confidence interval
construction), one can in principle use Lemma \ref{LemmaVar}, replacing the
unknown quantities with their empirical counterparts. However, this approach
does not work well for the real data that we have tried for two reasons: (a)
\textit{Multiple} integrals on empirical third or higher moments tend to be
highly variable. (b) The computing demands are quite high given the many
multiple integrals involved. Fortunately, we find an efficient shortcut.
First, when the mean arrival rate $\mu $ is large, variation from the
underlying stochastic arrival rate dominates in the variance decomposition (\ref{vardec}):
$\operatorname{var}\{E(\hat{C}_{\mu ,h}(t)|\lambda (\cdot))\}\gg
E\{\operatorname{var}(\hat{C}_{\mu ,h}(t)|\lambda (\cdot))\}$. Proposition \ref{varRatio}
 below gives a theoretical justification. Second, for the real
biophysical experimental data that we have tried, $\operatorname{var}\{E(\hat{C}
_{\mu ,h}(t)|\lambda (\cdot))\}$ accounts for more than 95\% of the total
variance $\operatorname{var}(\hat{C}_{\mu ,h}(t))$. Furthermore, in these real
data, $A_{t,T}^h/(T-2bh-t)^{2}$ in the decomposition (\ref{var1}) of
$\operatorname{var}\{E(\hat{C}_{\mu ,h}(t)|\lambda (\cdot))\}$ provides more than
90\% of $\operatorname{var}(\hat{C}_{\mu ,h}(t))$. These observations suggest that
we can use $A_{t,T}^h/(T-2bh-t)^{2}$ to approximate $\operatorname{var}(\hat{C}
_{\mu ,h}(t))$.

\begin{proposition}\label{varRatioA}
\label{varRatio}Denote $\lambda _{0}(t)=\lambda (t)/\mu $, that is, $E(\lambda
_{0}(t))=1$ and $\lambda (t)=\mu \lambda _{0}(t)$. Suppose the law of $
\{\lambda _{0}(t),t\in \mathsf{R}\}$ is fixed. Then under Assumptions \ref{ass1}--\ref{ass3},
for any fixed $T$, $h$ and $t$,
\begin{equation}
\frac{A_{t,T}^h}{(T-2bh-t)^{2}}\big/\operatorname{var}(\hat{C}_{\mu ,h}(t))\rightarrow
1,\qquad\mbox{as }\mu \rightarrow \infty,
\end{equation}
where $A_{t,T}^h$ is defined in Lemma \ref{LemmaVar}.
\end{proposition}

This proposition directly relates to real experimental data, especially
those from fluorescence biophysical experiments. In such experiments the
samples are usually placed under a laser beam, and the photon arrival
intensity is proportional to the laser strength. To illuminate the sample,
experimenters usually use a strong laser. In this scenario, since the
intrinsic molecular dynamics do not change, the law of $\{\lambda _{0}(t)\}$
remains the same, while $\mu $ is large.

The approximation of $A_{t,T}^h/(T-2bh-t)^{2}$ can be further simplified for
practical use. First, because the bandwidth $h$ is usually chosen to be
small, and $v(\cdot ,\cdot ,\cdot)$ is a continuous function, $A_{t,T}^h$
approximately equals $\int \int_{[bh,T-bh-t]^{2}}\*v(t,t,s^{\prime
}-s)\,ds\,ds^{\prime }$. Second, since the process $\{\lambda (s),s\in \mathsf{R}
\}$ is stationary, and to accurately estimate $C(t)$, $t$ is usually small
compared to $T$ (in order to have enough data), $\int
\int_{[bh,T-bh-t]^{2}}v(t,t,s^{\prime }-s)\,ds\,ds^{\prime }$ approximately
equals $\frac{2}{(T-2bh-t)^{2}}\!\int_{0}^{T-t-2bh}(T-t-r)v(t,t,r)\,dr$. Third,
replacing
\[
v(t,t,r)=E\bigl( \bigl(\lambda (0)-\mu \bigr)\bigl(\lambda (t)-\mu \bigr)\bigl(\lambda (r)-\mu
\bigr)\bigl(\lambda (r+t)-\mu \bigr)\bigr) -C^{2}(t)
\]
with its empirical counterpart $\operatorname{\hat{cov}}(t,r)$, which is
\begin{eqnarray*}\label{varEst}
&&\operatorname{\hat{cov}}(t,r) =\max \biggl\{\frac{1}{T-2bh-r-t}\\
&&\hphantom{\operatorname{\hat{cov}}(t,r) =\max \biggl\{}{}\times\int_{bh}^{T-bh-r-t}\bigl(\hat{\lambda
}_{h}(s)-\hat{\mu}\bigr)\bigl(\hat{\lambda}_{h}(s+t)-\hat{\mu}\bigr) \bigl(\hat{\lambda}_{h}(r+s)-\hat{\mu}\bigr)\\
&&\hspace{21pt}\hphantom{\hat{cov}(t,r) =\max \biggl\{{}\times\int_{bh}^{T-bh-r-t}}{}\times\bigl(\hat{\lambda}_{h}(r+s+t)-\hat{\mu}\bigr)\,ds-\hat{C}_{\hat{\mu},h}^{2}(t),0\biggr\},
\end{eqnarray*}
results in the final approximation of $\operatorname{var}(\hat{C}_{\mu ,h}(t))$:
\begin{equation}
\hat{V}(t)=\frac{2}{(T-2bh-t)^{2}}\int_{0}^{T-t-2bh}(T-t-2bh-r)\operatorname{\hat{cov}}
(t,r)\ dr.
\end{equation}
Note that $v(t,t,r)$ is typically nonnegative, so we force $\operatorname{\hat{cov}}(t,r)$
to be nonnegative also. We will demonstrate the use of $\hat{V}(t)$ in
Section \ref{Numerical}.

\subsubsection*{Comparison with existing methods}
\begin{figure}

\includegraphics{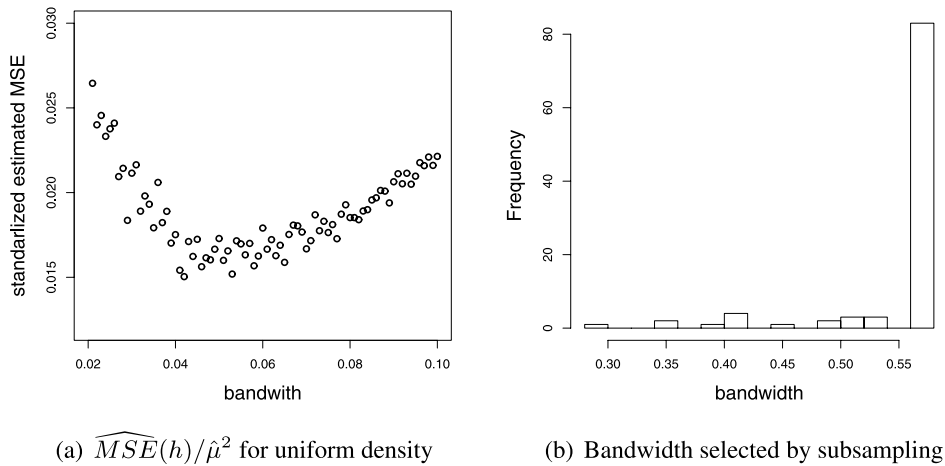}

\caption{\textup{(a)} $\widehat{\operatorname{MSE}}(h)$ vs. bandwidth using
Diggle's (\protect\citeyear{Diggle1985}) method
for data from the two-state model. \textup{(b)} The bandwidth selected by the
subsampling procedure. The histogram is based on 100 i.i.d. replications of
the Cox process from the two-state model.}\label{MSEfig}
\end{figure}

 Diggle (\citeyear{Diggle1985}) provided a
procedure for selecting the bandwidth for estimating the arrival rate $
\lambda (t)$ in the case of $f$ being the uniform kernel. In this procedure,
based on an estimate $\widehat{\mathit{MSE}}(h)$ of $E(\hat{\lambda}_{h}(t)-\lambda
(t))^{2},$ the bandwidth is chosen to be the one that gives the smallest $
\widehat{\mathit{MSE}}(h)$. Figure \ref{MSEfig}(a) shows the standardized estimate $
\widehat{\mathit{MSE}}(h)/\hat{\mu}^{2}$ for the data of Figure~\ref{fig1}(a) by this approach. However, this method is
computationally more intensive than our method, since it involves estimating
MSE for all the possible bandwidths. Moreover, the MSE estimator provided by
\citet{Diggle1985} is only meant for the uniform kernel and does not generalize
to other kernels.

\citet{Guan2007} has proposed a composite likelihood cross-validation approach in
selecting bandwidth for estimating the ACF. However, due to the large data
size in our study (more than two million arrival points), this method is
computationally too expensive to use (we found that the C program cannot
even finish in an affordable time).

We also applied the subsampling procedure of Guan, Sherman and Calvin (\citeyear{GuanShe2004}, \citeyear{GuanShe2006})
to our data. Figure \ref{MSEfig}(b) shows the histogram of the
bandwidths selected by the subsampling procedure based on 100
i.i.d. simulations from the two-state model in Section \ref{sec2}. We found that this
procedure leads to a much larger bandwidth than $\hat{h}_{\mathrm{opt}}$ proposed in
Section \ref{sec2}, and, consequently, the estimates of $C(t)$ have large bias,
particularly for $t$ close to zero.

Compared to the existing methods, in terms of computational effort, our
proposed method takes no more than five minutes to finish analyzing a
process with more than two million data points, including estimating the
arrival rate and the autocorrelation function and constructing the
confidence intervals.

\subsubsection*{Connection with the classical kernel density estimate} Despite\vspace{1pt} its
similarity with the classical kernel density estimate, the kernel estimates $
\hat{\lambda}_{h}(t)$ and $\hat{C}_{\mu,h}(t)$ have several distinct
features: (i) Since $\lambda (t)$ is stochastic, a consistent estimate of $
\lambda (t)$ does not exist. (ii) In classical kernel problems, the number
of observations $K$ does not depend on the underlying density, whereas the
total number of observations in our case is random and depends on the
stochastic process $\{\lambda (t),t\in \lbrack 0,T]\}$. Consequently, (iii)
consistency refers to the observational window $T\rightarrow \infty $. (iv)~The asymptotic behavior of the kernel estimate would depend on the
distributional properties of $\lambda (t)$, as we shall see next.

\section{Asymptotic distribution of the kernel estimate}\label{sec4}

We investigate the limiting behavior of the kernel estimate $
\hat{C}_{\mu ,h}(t)$ in this section, since the asymptotic normality plays
an important role in confidence interval construction. For well-behaved $
\lambda (t)$, we can show that the asymptotic normality of $\hat{C}_{\mu
,h}(t)$ holds.

\subsection*{$\rho$-mixing arrival rate} Let $\mathcal{F}_{t}=\sigma
(\lambda (s)\dvtx s\leq t)$ be the sigma field generated by $\lambda (s)$ for $
s\leq t$, and $\mathcal{G}_{t}=\sigma (\lambda (s)\dvtx s\geq t)$ be the tail
sigma field generated by $\lambda (s)$ for $s\geq t$. Define
\begin{equation} \label{rhot}
\qquad\rho _{t}=\sup \{E(\xi \eta)\dvtx\xi \in \mathcal{F}_{s},E\xi =0,\Vert \xi
\Vert \leq 1;\eta \in \mathcal{G}_{s+t},E\eta =0,\Vert \eta
\Vert \leq 1\}.
\end{equation}
$\lambda (t)$ is said to be finite $\rho $-mixing if $\int_{0}^{\infty }\rho
_{s}\,ds<\infty $ [\citet{Billing1999}].

\begin{theorem}\label{Cor1}
Suppose that Assumptions \ref{ass1}, \ref{ass2}, \ref{ass4} and \ref{ass5} hold, and that the
arrival rate process $\{\lambda (t),t\in \mathsf{R}\}$ is bounded and finite
$\rho $-mixing. Then for fixed $t,h\geq 0$,
\begin{equation} \label{limit}
\sqrt{T}[\hat{C}_{\mu ,h}(t)-E(\hat{C}_{\mu ,h}(t))]\stackrel{D}{\rightarrow}
N(0,\sigma ^{2}(t,h))\qquad\mbox{as }T\rightarrow \infty ,
\end{equation}
where $\sigma ^{2}(t,h)=\lim_{T\rightarrow \infty }T\operatorname{var}(\hat{C}
_{\mu ,h}(t))$.
\end{theorem}

\begin{theorem}\label{CLT4}
Suppose that Assumptions \ref{ass1} and \ref{ass2} hold and that the stochastic
process $\lambda (s)$ is a continuous-time Markov chain with finite number
of states. Then for fixed $t,h\geq 0$, the asymptotic normality (\ref{limit})
holds with $\sigma ^{2}(t,h)=\lim_{T\rightarrow \infty }T\operatorname{var}
(\hat{C}_{\mu ,h}(t))$.
\end{theorem}

Theorem \ref{CLT4} covers a large class of arrival rates. Another class of
processes, widely used in the physical science literature, is functionals of
stationary Gaussian processes: $\lambda (s)=g(W(s))$, where $g$ is a
positive and continuous function, and $\{W(s),s\in \mathsf{R}\}$ is a
zero-mean Gaussian process. We will see that as long as the autocorrelation
of $W(s)$ decays reasonably fast, the asymptotic normality of $\hat{C}_{\mu
,h}(t)$ remains true.

We consider, in particular, Gaussian processes of the form
\begin{equation}  \label{disGauss}
\qquad\bigl\{W(t),t\in \mathsf{R}\dvtx W(t)=W(j\varepsilon)\mbox{ for }t\in\bigl[
j\varepsilon ,(j+1)\varepsilon\bigr),j=0,\pm 1,\pm 2,\ldots \bigr\},
\end{equation}
where $\varepsilon >0$ is a fixed constant. In other words, we consider
Gaussian processes generated piecewise by a discrete skeleton: $\{\ldots
,W(-2\varepsilon ),W(-\varepsilon),\break W(0),W(\varepsilon), W(2\varepsilon
),\ldots \}$, where $W(j\varepsilon)$ is a discrete-time stationary
zero-mean Gaussian process. The reason that we focus on this type of
Gaussian process is two fold: first, if $\varepsilon $ is small enough, $W$
can essentially approximate any continuous stationary Gaussian process with
arbitrary precision, and this is typically how one simulates a Gaussian
process; second, the theoretical calculations behind continuous-time
Gaussian processes, especially those regarding mixing conditions, are quite
delicate [see, e.g., \citet{IbrRoz1978}], so to avoid
drifting too much into the mathematical details and to present our proofs in
a concise manner, we work on (\ref{disGauss}). We have the following result
on functionals of Gaussian processes; its proof is given in the
supplementary material  [\citet{ZhaKou2010}].

\begin{theorem}\label{DiscreteGaussian}
 Suppose that Assumptions \ref{ass1}, \ref{ass2}, \ref{ass4} and \ref{ass5} hold, and
that $\lambda (s)=g(W(s))$, where $g$ is a positive and bounded measurable
function, and $\{W(s),s\in \mathsf{R}\}$, defined in (\ref{disGauss}), is
generated from a discrete skeleton $\{W(j\varepsilon)\}$. If the ACF $
\gamma (j\varepsilon)=\operatorname{cov}(W(0),W(j\varepsilon))$ satisfies $
\sum_{j=0}^{\infty }|\gamma (j\varepsilon)|<\infty $, then for any fixed $t$
and $h$, the asymptotic normality (\ref{limit}) of $\hat{C}_{\mu ,h}(t)$
holds with $\sigma ^{2}(t,h)=\lim_{T\rightarrow \infty }T\operatorname{var}(\hat{C}_{\mu ,h}(t))$.
\end{theorem}

\subsection*{Long-range dependent processes} Stochastic processes with
a finite integrated correlation $\int_{0}^{\infty }C(t)\,dt<\infty $ are said
to be short-range dependent. Our results essentially say that for Cox
processes with short-range dependent arrival rates, we expect the asymptotic
normality of $\hat{C}_{\mu ,h}(t)$ to hold, which offers a big advantage in
the confidence interval construction. For long-range dependent arrival rates
($\int_{0}^{\infty }C(t)\,dt=\infty $), however, no easy conclusion can be
drawn about the asymptotic behavior of $\hat{C}_{\mu ,h}(t)$. Even the form
of limiting law varies from case to case. For example, the limiting process
might be a fractional Brownian Motion [\citet{Whitt2002}], a stable L$\mathrm{\acute{e}}$vy motion
[\citet{Whitt2002}] or a Rosenblatt process [\citet{Taqqu1975}]. Moreover, the variance
of the limiting law, most likely, will not be the same as the limit of the
variance [\citet{Taqqu1975}], that is, $\lim_{T\rightarrow \infty }\operatorname{var}(\hat{
C}_{\mu ,h}(t)/\sqrt{\operatorname{var}(\hat{C}_{\mu ,h}(t))})\neq \operatorname{var}
(\lim_{T\rightarrow \infty }\hat{C}_{\mu ,h}(t)/\sqrt{\operatorname{var}(\hat{C}_{\mu ,h}(t))})$. Therefore, an interesting open problem is to investigate
the asymptotic behavior of the Cox process estimates with long-range
dependent arrival rates.

\section{Numerical study of ACF estimation}\label{Numerical}
We illustrate our method through several numerical examples---both simulation and real biophysical experimental data. We use the
Epanechnikov kernel throughout this section.

\subsection{Simulation examples}
\subsubsection*{Finite-state Markov chains} Since Theorem \ref{CLT4} guarantees the
asymptotic normality of the kernel ACF estimate, we can construct a
pointwise $1-\alpha $ approximate confidence interval (C.I.) of $C(t)$ via
\begin{equation}\label{CI}
\biggl[\tilde{C}_{\hat{\mu},h}(t)-\Phi ^{-1}\biggl(1-\frac{\alpha }{2}\biggr)\sqrt{\hat{V}(t)},
\tilde{C}_{\hat{\mu},h}(t)+\Phi ^{-1}\biggl(1-\frac{\alpha }{2}\biggr)\sqrt{\hat{V}(t)}\biggr],
\end{equation}
where $\tilde{C}_{\hat{\mu},h}(t)$ and $\hat{V}(t)$ are given in (\ref{BiasCorrectCt})
and (\ref{varEst}) respectively. Following the discussion
in Section \ref{Sec3.2}, we use $h=\min (5/\hat{\mu},\hat{h}_{\mathrm{opt}})$ for $
t<2b\hat{h}_{\mathrm{opt}}$ and $h=\hat{h}_{\mathrm{opt}}$ for $t\geq 2b\hat{h}_{\mathrm{opt}}$.

\begin{figure}[b]

\includegraphics{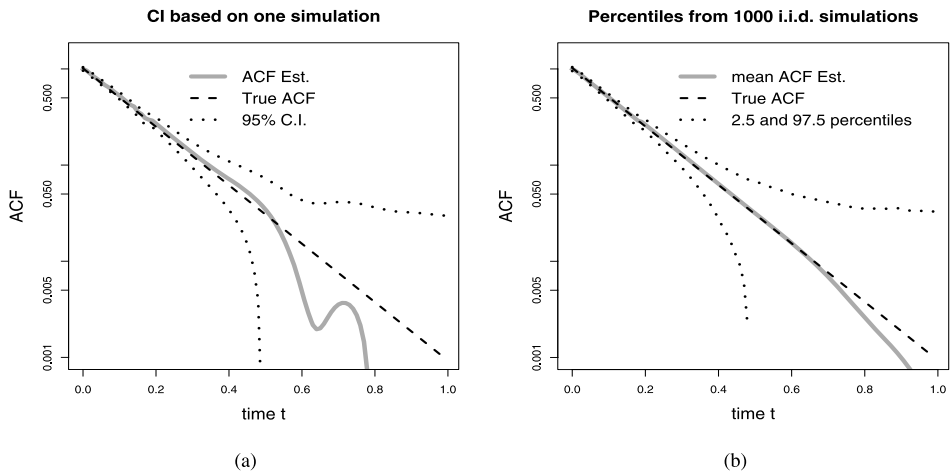}

\caption{ACF estimation for two-state Markov chains. The left panel shows $
\tilde{C}_{\hat{\mu},h}(t)$ (the time $t$ is in second) and the
approximate 95\% C.I. [normalized by $\tilde{C}_{\hat{\mu},h}(0)$]
based on one sequence of arrival data. The right panel shows the 2.5 and
97.5 percentiles of $\tilde{C}_{\hat{\mu},h}(t)$ calculated from
1000 i.i.d. repetitions from the same model.}\label{TwoState}
\end{figure}

\begin{figure}[b]

\includegraphics{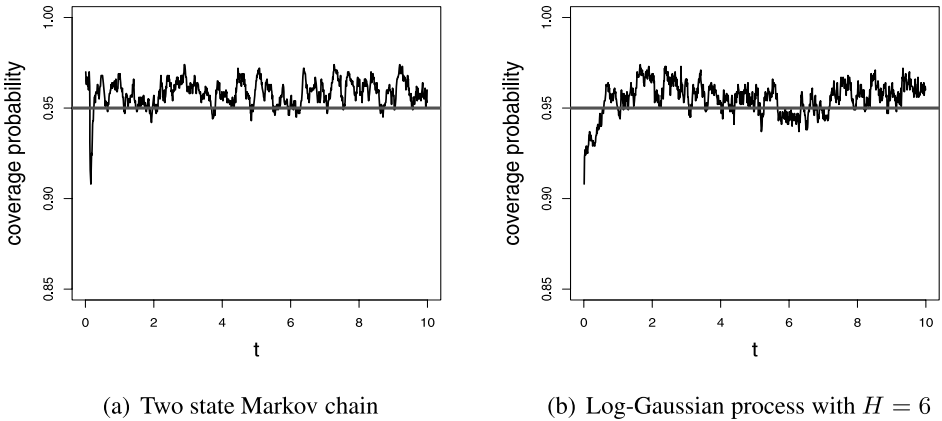}

\caption{The coverage probabilities of the 95\% C.I. (\protect\ref{CI}) for $
t\in[0,10]$ (the time $t$ is in second) under different models.}\label{ProbCov}
\end{figure}

We revisit the two-state Markov chain model (\ref{2state}) in Section
\ref{sec2}. In this case, the true ACF is exponential: $C(t)=(\lambda
_{A}-\lambda _{B})^{2}k_{1}k_{2}\exp (-(k_{1}+k_{2})t)/\break (k_{1}+k_{2})^{2}$.
We applied the kernel estimator and (\ref{CI}) to the data set simulated in
Section \ref{sec2} [Figure~\ref{fig1}(a)]. Figure~\ref{TwoState}(a)
shows $\tilde{C}_{\hat{\mu},h}(t)$ as the solid line, and the point-wise
95\% C.I. as the dotted lines. The true ACF $C(t)$, shown as the dashed
line, is well recovered. Since $C(t)$ usually decays quite fast, to
highlight the details, especially around the tails, we plotted the estimate
on the logarithm scale. We see from Figure~\ref{TwoState}(a) that $\log \tilde{C}_{\hat{\mu},h}(t)$ is linear with $t$,
 indicating the exponential decay of
$C(t)$. As a check for the accuracy of the C.I., we repeated the data
generation 1000 times independently. For each simulated data set, we
calculated $\tilde{C}_{\hat{\mu},h}(t)$. The 2.5 and 97.5 percentiles of
these repeated estimates $\tilde{C}_{\hat{\mu},h}(t)$ give the real 95\%
coverage of $\tilde{C}_{\hat{\mu},h}(t)$, which is shown on Figure \ref{TwoState}(b). Comparing the two panels, we see that the variance estimate
based on just one realization is close to the truth. From the 1000
i.i.d. repetitions, we calculated the coverage probabilities of the 95\%
C.I. (\ref{CI}) for various $t$. Table \ref{tab3} (the second row) reports
the numbers, which are close to the nominal 95\%; Figure \ref{ProbCov}(a) plots them graphically.

\begin{table}
\tabcolsep=4pt
\caption{The coverage probabilities of the 95\% C.I. (\protect\ref{CI}) at
various time points for different models based on 1000 i.i.d. repetitions.
For reference, the standard deviation of a binomial proportion with success
probability of 0.95 and 1000 trails is 0.0069}\label{tab3}
\begin{tabular*}{\textwidth}{@{\extracolsep{\fill}}lcccccccccc@{}}
\hline
 & \multicolumn{10}{c@{}}{\textbf{Time} $\bolds{t}$}\\ [-7pt]
&\multicolumn{10}{c@{}}{\hrulefill}\\
\multirow{2}{45pt}[10pt]{\textbf{Coverage probability}} & \textbf{0.01} & \textbf{0.02} & \textbf{0.05} & \textbf{0.1} & \textbf{0.2}
& \textbf{0.5} & \textbf{1} & \textbf{2} & \textbf{5} & \textbf{10}
\\
\hline
Two state & 0.97 & 0.97 & 0.97 & 0.96 & 0.93 & 0.97 & 0.97 & 0.96 & 0.97 &
0.96 \\
Log-Gaussian $H=6$ & 0.91 & 0.92 & 0.92 & 0.93 & 0.93 & 0.94 & 0.95 & 0.97 &
0.95 & 0.96 \\
Log-Gaussian $H=0.5$ & 0.57 & 0.58 & 0.59 & 0.60 & 0.62 & 0.63 & 0.66 & 0.66
& 0.65 & 0.66\\
\hline
\end{tabular*}
\end{table}

\begin{figure}

\includegraphics{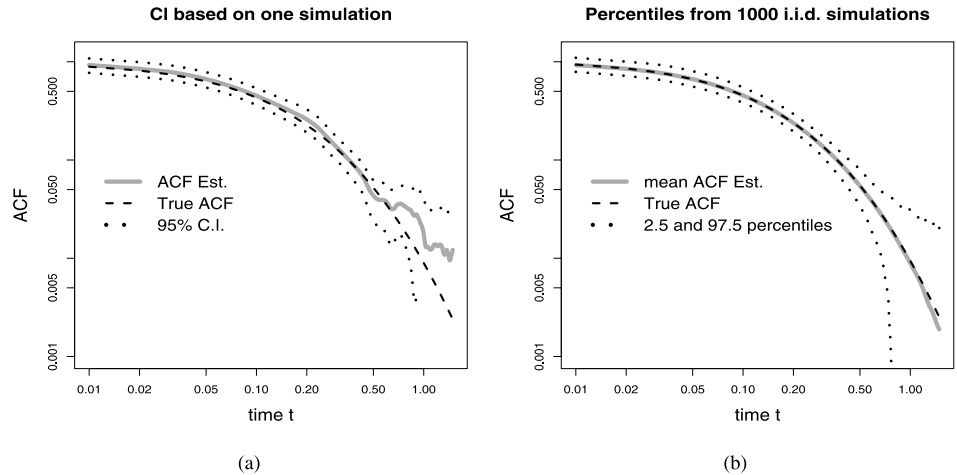}

\caption{ACF estimation for a short-range dependent log-Gaussian process
with $C(t)=10^{6}(\exp (1/(1+|t|)^{6}+1)-e)$ (the time $t$ is in second).
The left panel shows $\tilde{C}_{\hat{\mu},h}(t)$ and the
approximate 95\% C.I. [normalized by $\tilde{C}_{\hat{\mu},h}(0)$]
based on one sequence of arrival data. The right panel shows the 2.5 and
97.5 percentiles of $\tilde{C}_{\hat{\mu},h}(t)$ calculated from
1000 i.i.d. replications from the same model. The total observational time $
T $ equals $1500$. Both graphs are plotted on the log--log scale.}\label{ACFLog1}
\end{figure}

\subsubsection*{Log Gaussian processes}We next consider examples where
the arrival rate $\lambda (t)$ follows a log Gaussian process: $\lambda
(t)=M\exp \{W(t)\}$, where $W(t)$ is a stationary zero-mean Gaussian
process with the ACF $\gamma (t)$. As we mentioned in Section~\ref{sec2},
the log Gaussian process has been used to model the conformational dynamics
and reactivity of enzyme molecules [\citet{Min2005a}; \citet{KouXie2004}]. As in Section
\ref{sec2}, we take $\gamma (t)=1/(1+a|t|)^{H}$ so that $
C(t)=M^{2}(\exp (\gamma (t)+1)-e)$ decreases at the order of $t^{-H}$. Both $
a$ and $H$ are positive constants. The larger the decay slope $H$, the
faster the $C(t)$ converges to zero and the faster the estimate converges to
$C(t)$. $H$ also determines the dependence structure of the Cox process: if $
H\leq 1$, the process is long-range dependent. In the simulation, we
generate $W(t)$ through the discrete skeleton (\ref{disGauss}), and then
draw the arrival times $s_{1},s_{2},\ldots $ on top of $\lambda (t)$. For
each simulated arrival sequence, we calculate the kernel estimate $\tilde{C}
_{\hat{\mu},h}(t)$ and the 95\% C.I. (\ref{CI}).

We consider two log Gaussian processes: one with $H=6$ and $a=1$, and the
other with $H=0.5$ and $a=20$. In both cases, the maximum observational time
$T=1500$ and the constant $M$ is taken to be $1000$ to mimic typical photon
arrival data from a biophysical experiment. For the log Gaussian process
with $H=6$, Figure \ref{ACFLog1}(a) plots $\tilde{C}_{\hat{\mu},h}(t)$ and
the 95\% approximate C.I. based on one data set. Figure \ref{ACFLog1}(b)
plots the 2.5 and 97.5 percentiles of $\tilde{C}_{\hat{\mu},h}(t)$ from 1000
i.i.d. repetitions. For easy visual detection of the power law decay, the
graph is plotted on a log--log scale. The similarity between the left and
right panels indicates the effectiveness of our method. The real coverage
probabilities of the 95\% C.I. (\ref{CI}) are shown in Figure \ref{ProbCov}(b) and Table \ref{tab3} (the third row). We see that the real
coverage probabilities for the $H=6$ case are close to the nominal 95\%.
Figure \ref{ACFLog1} also suggests that $\log C(t)$ is roughly linear with $
\log t$ near the tail of the curve.

\begin{figure}

\includegraphics{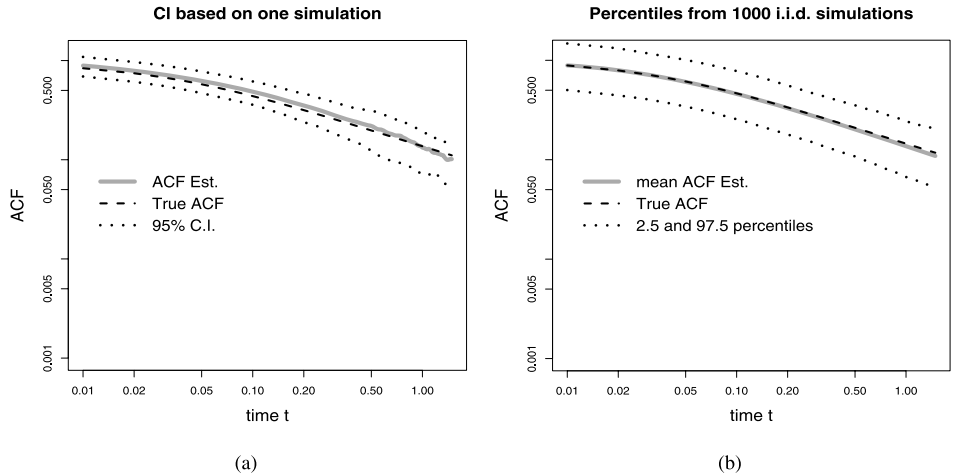}

\caption{ACF estimation for a long-range dependent log-Gaussian process with
$C(t)=10^{6}(\exp (1/(1+20|t|)^{0.5}+1)-e)$ (the time $t$ is in second). The
left panel shows $\tilde{C}_{\hat{\mu},h}(t)$ and the approximate
95\% C.I. [normalized by $\tilde{C}_{\hat{\mu},h}(0)$] based on one
sequence of arrival data. The right panel shows the 2.5 and 97.5 percentiles
of $\tilde{C}_{\hat{\mu},h}(t)$ calculated from 1000
i.i.d. replications from the same model. The total observational time $T$
equals $1500$. Both graphs are plotted on the log--log scale.}\label{ACFLog}
\end{figure}

The log Gaussian process with $H=0.5$ and $a=20$ is long-range dependent. We
can still calculate $\tilde{C}_{\hat{\mu},h}(t)$ and the C.I. (\ref{CI}).
However, since even asymptotic normality is no longer valid, one would
expect the real coverage to be way off. Figure \ref{ACFLog} contrasts the
``C.I.'' based on one data set with the
true 2.5 and 97.5 percentiles of $\tilde{C}_{\hat{\mu},h}(t)$ from 1000
i.i.d. replications. It is evident that although $\tilde{C}_{\hat{\mu},h}(t)$
still estimates $C(t)$ reasonably well, the ``C.I.'' constructed from one data set is quite narrower than
the true percentiles. The last row of Table \ref{tab3} shows that the real
coverage probabilities of (\ref{CI}) in this long-range dependent case are
much smaller than the nominal 95\%---clearly the asymptotic variance is
underestimated.

\subsubsection*{Static processes} 
\begin{figure}

\includegraphics{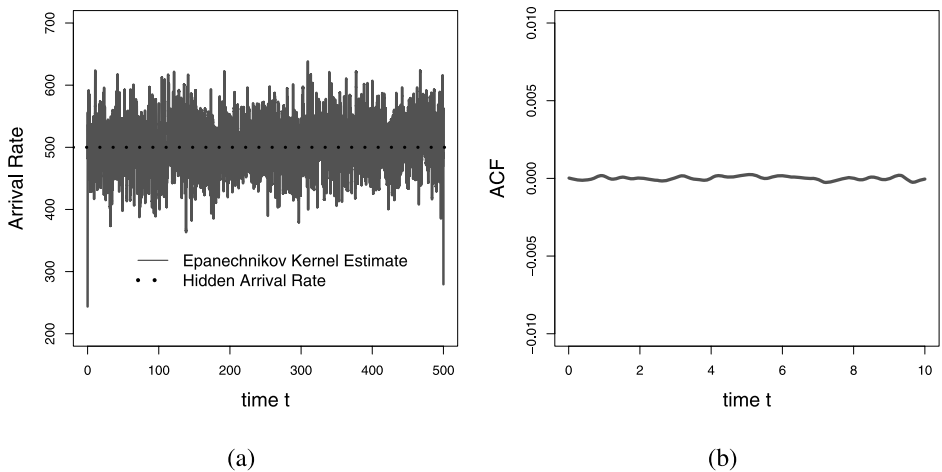}

\caption{Analyzing Cox process data with constant arrival rate $\lambda (t)\equiv 500$
and $T=500$.
\textup{(a)} Arrival rate estimate. \textup{(b)} ACF estimate (normalized by $\hat{\mu}^{2}$).}\label{ConstRate}
\end{figure}

Our method can be easily applied to
detect static processes. When the underlying biological process is static,
the photon arrival rate $\lambda (t)$ is a constant, and $C(t)\equiv 0$ for
$t>0$. In this case, we would observe that the arrival rate estimate
$\hat{\lambda}_{\hat{h}_{\mathrm{opt}}}(t)$ oscillates around a constant, and the ACF
estimate $\tilde{C}_{\hat{\mu},h}(t)$ clusters around zero. Figure \ref{ConstRate}
shows such an example with constant arrival rate $\lambda(t)\equiv 500$ and $T=500$.

\subsection{Experimental photon arrival data}

Studying the conformational dynamics of proteins is of current biophysical
interest. For example, scientists have become aware that an enzyme's
conformational fluctuation can directly affect its catalytic activity---certain
 conformations yield highly active catalysis, whereas others lead to
less active catalysis [\citet{LuXun1998}; \citet{English2006}].
 A~recent single-molecule experiment [\citet{Yang2003}]
investigates the conformational dynamics of a protein-enzyme compound
\textit{Fre}, which is involved in the DNA synthesis of \textit{E. Coli}.
In the experiment, the protein compound is immobilized and placed under a
laser beam. Photons from the laser-excited molecule are collected. Since the
photon arrival rate depends on the molecule's time-varying three-dimensional
conformation (different conformations of \textit{Fre} generate different
arrival rates), the spontaneous conformation fluctuation of \textit{Fre}
leads to a stochastic arrival rate. The ACF of the photon arrival rates
therefore reflects the time dependence of \textit{Fre}'s conformational
fluctuation [\citet{Weiss2000}].

\begin{figure}

\includegraphics{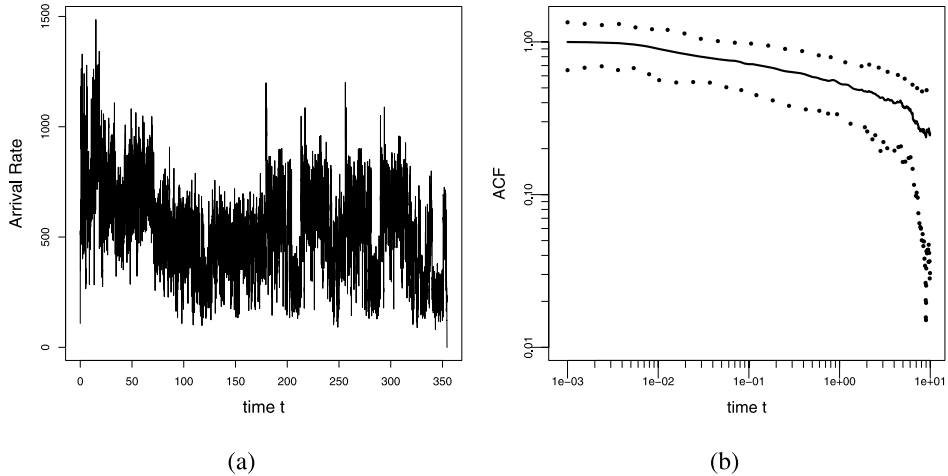}

\caption{Analyzing the photon arrival data from a single-molecule
experiment with $T=354.3$ and $\hat\mu=534.6$. \textup{(a)} Arrival rate
estimate. \textup{(b)} ACF estimate and its 95\% confidence interval plotted on the
log--log scale.}\label{SmallData}
\end{figure}

The experimental photon arrival data has a total observational time $T=354$
seconds. The empirical mean arrival rate $\hat{\mu}=534.6$ counts/second. We
first estimated the arrival rate and showed it in Figure
\ref{SmallData}(a). This plot leads to a natural question regarding the nature of \textit{Fre}'s
 conformational fluctuation: does \textit{Fre} have a small number of
distinct conformation states or many? Looking at the decay of $C(t)$
provides one way to address this question. We applied the bias-corrected $
\tilde{C}_{\hat{\mu},h}(t)$ to estimate $C(t)$. Figure \ref{SmallData}(b)
shows $\tilde{C}_{\hat{\mu},,h}(t)$ and its approximate 95\% C.I. (\ref{CI}). We plotted the estimates on a log--log scale to give a better view of the
decay of the ACF. The apparent linear pattern suggests a power-law
relationship. If there are only two, three or even four conformation states,
then $C(t)$ should be a mixture of no more than three exponential functions.
The apparent power-law relationship indicates a different picture: instead
of having two, three or even four discrete conformation states, the 3D
conformation of \textit{Fre} appears to fluctuate over a continuum (as a
check, we have attempted to parametrically fit a mixture of three
exponentials to the estimated ACF, but even the best fitting is very poor),
so the parametric finite state Markov Chain model cannot be applied here.
The slow decay of the ACF thus points to a complicated conformation dynamic
of \textit{Fre}, which implies that the enzyme's catalytic rate could vary
over a broad range, a phenomenon called dynamic disorder in the biophysics
literature [\citet{Min2005a}; \citet{Lerch2005}].

\begin{figure}

\includegraphics{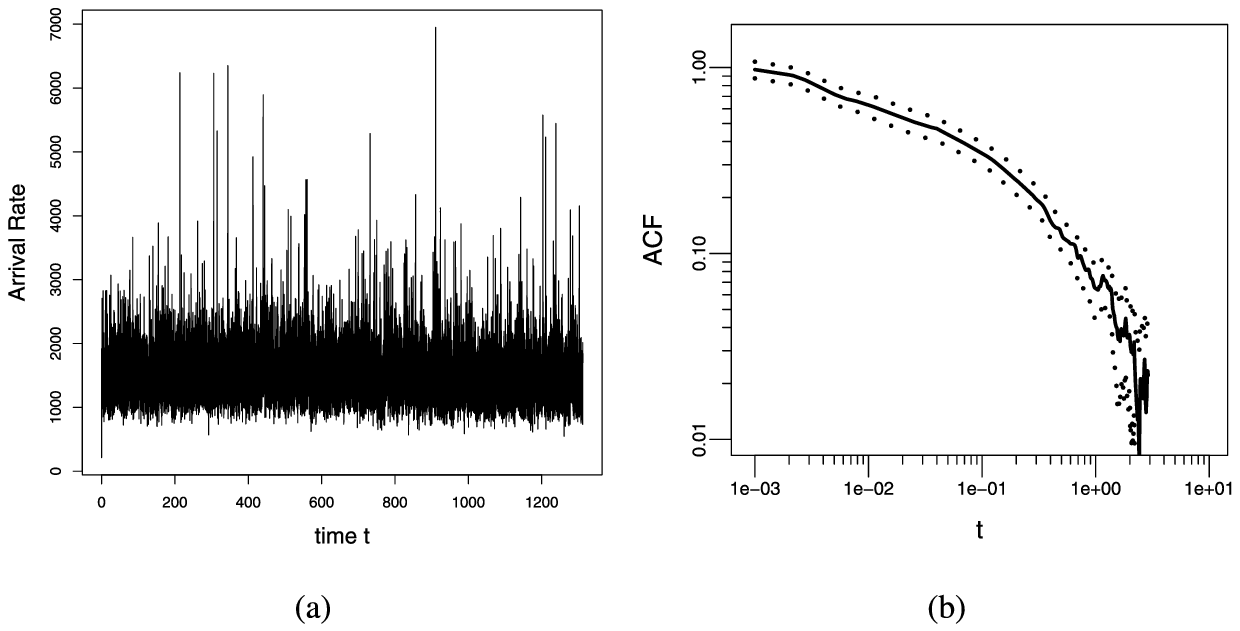}

\caption{Analyzing the photon arrival data from another single-molecule
experiment with $T=1312.8$ and $\hat\mu=1523.5$. \textup{(a)} Arrival rate
estimate. \textup{(b)} ACF estimate and its 95\% confidence interval plotted on the
log--log scale.}\label{LargeData}
\end{figure}

Another recent single-molecule experiment [\citet{Min2005b}] also
investigates protein's conformational dynamics, studying a protein complex
formed by fluorescein and monoclonal antifluorescein. This protein complex
is an antibody-antigen system. Like the previous compound, the 3D
conformation of the molecule spontaneously fluctuates over time. To study
the conformational dynamics, the immobilized protein complex was placed
under a laser beam. Photons from the laser-excited molecule are collected.
The photon arrival rate $\lambda (t)$ depends \mbox{on the}~molecule's time-varying
conformation. Figure \ref{LargeData}(a) shows the arrival rate estimates
for this data, which have $T=1312.8$ and $\hat{\mu}=1523.5$. This plot seems
to suggest that there are many conformation states in this antibody-antigen
system. To further investigate, we applied $\tilde{C}_{\hat{\mu},h}(t)$ to
estimate $C(t)$. Figure \ref{LargeData}(b) shows our estimate and the
approximate 95\% C.I. (\ref{CI}) on a log--log scale. Again, we observed a
slow decay of the ACF. We attempted to parametrically fit a mixture of three
exponentials to the estimated ACF but only obtained a very poor result. Like
the previous system, it appears that the 3D conformation of this
antibody-antigen system fluctuates over a broad range rather than over just
a few, say, three of four, discrete states.

Since the second experiment is on a totally different system from the first,
our statistical results indicate that (i) conformational fluctuation could
be widely present in protein systems; (ii) the fluctuation appears to be
over a broad range of time scales. Our results thus support the growing
understanding in the biophysics community that proteins' conformational
fluctuation is a complex phenomenon, which in turn affects some crucial
functions of proteins, such as enzyme catalysis [\citet{Lerch2005};
\citet{English2006}] and electron transfer in photosynthesis
[\citet{Wang2007}].

\section{Conclusion}\label{sec6}

Motivated by the analysis of experimental data from biophysics, we propose a
nonparametric kernel based method for inferring Cox process in this article,
complementing existing parametric approaches. An important feature of the
arrival data in biophysics is that the arrival rate is often large, which
makes the methods developed for analyzing spatial point processes (which
usually have low arrival rates), such as variance estimate, bandwidth
selection and asymptotic theory, not directly applicable for our purpose. In
addition to proposing the kernel estimates, we conduct a detailed study of
their properties. We show that the asymptotic normality of our ACF estimates
holds for most short-range dependent processes, which provides the
theoretical underpinning for confidence interval construction. We provide an
approximation of the variance of the ACF estimate, which accounts for at
least 90\% of the total variation in our examples. We can possibly improve
this approximation, for example, by taking into account the Poisson
variation $E\{\operatorname{var}(\hat{C}_{\mu ,h}(t)|\lambda (\cdot))\}$ part, which might
be particularly beneficial when $\lambda (t)$ has a strong short-range
dependence.

We applied our nonparametric method to analyze two real photon arrival data
produced in recent (single-molecule) biophysical experiments. Using our
kernel ACF estimate, we examine the conformational dynamics of two different
protein systems. We observed that the conformational fluctuation exhibits a
long memory and spans a broad range of time scales, confirming the recent
experimental discovery that the classical static picture of proteins that
researchers used to assume needs to be revised.

An important open question for future study is to investigate Cox processes
with long-range dependent arrival rates. Another open question for our
future investigation is the estimation of high-order correlations of the
arrival rate, as biophysicists and chemists have used them to discriminate
different mechanistic and phenomenological models [\citet{Mukamel1995}].

\textbf{Acknowledgments.} The authors are grateful to the Xie group at the
Department of Chemistry and Chemical Biology of Harvard University for
sharing the experimental data. We also thank Yongtao Guan, the editor, the
associate editor and the referees for helpful suggestions.

\begin{supplement}[id=suppA]
\stitle{Technical proofs}
\slink[doi]{10.1214/10-AOAS352SUPP}
\slink[url]{http://lib.stat.cmu.edu/aoas/352/supplement.pdf}
\sdatatype{.pdf}
\sdescription{Technical proofs accompanying the paper ``Nonparametric inference of doubly
stochastic Poisson process data via the kernel method'' by Zhang and Kou.}
\end{supplement}

\printaddresses

\end{document}